\documentclass{aastex631}

\graphicspath{{./}{figures/}}

\usepackage[caption=false]{subfig}
\usepackage{multirow}

\begin{document}

\title{Classifying Seyfert galaxies with deep learning}

\correspondingauthor{Yen Chen Chen}
\email{yen-chen.chen@icranet.org}

\author[0000-0002-7543-2727]{Yen Chen Chen}
\affiliation{ICRA and Dipartimento di Fisica, Sapienza Universit\`{a} di Roma, Piazzale Aldo Moro 5, -00185 Rome, Italy}
\affiliation{ICRANet, Piazza della Repubblica 10, I-65122 Pescara, Italy}



\begin{abstract}

Traditional classification for subclass of the Seyfert galaxies is visual inspection or using a quantity defined as a flux ratio between the Balmer line and forbidden line. One algorithm of deep learning is Convolution Neural Network (CNN) and has shown successful classification results. We building a 1-dimension CNN model to distinguish Seyfert 1.9 spectra from Seyfert 2 galaxies. We find our model can recognize Seyfert 1.9 and Seyfert 2 spectra with an accuracy over 80\% and pick out an additional Seyfert 1.9 sample which was missed by visual inspection. We use the new Seyfert 1.9 sample to improve performance of our model and obtain a 91\% precision of Seyfert 1.9. These results indicate our model can pick out Seyfert 1.9 spectra among Seyfert 2 spectra. We decompose H$\alpha$ emission line of our Seyfert 1.9 galaxies by fitting 2 Gaussian components and derive line width and flux. We find velocity distribution of broad H$\alpha$ component of the new Seyfert 1.9 sample has an extending tail toward the higher end and luminosity of the new Seyfert 1.9 sample is slightly weaker than the original Seyfert 1.9 sample. This result indicates that our model can pick out the sources that have relatively weak broad H$\alpha$ component. Besides, we check distributions of the host galaxy morphology of our Seyfert 1.9 samples and find the distribution of the host galaxy morphology is dominant by large bulge galaxy. In the end, we present an online catalog of 1297 Seyfert 1.9 galaxies with measurement of H$\alpha$ emission line.

\end{abstract}



\section{Introduction} \label{sec:intro}

Active Galactic Nuclei (AGNs) are luminous sources across electromagnetic spectrum and show strong emission lines in optical spectra. The structures of an AGN are believed to consist of an accretion disk and a supermassive black hole embedded in an optically thick torus \citep{Rowan77,Antonucci85}. The AGN phenomenon is of accretion material near the central supermassive black hole, which releases a huge amount of energy ($L_\mathrm{bol} \approx 10^{48}~ \mathrm{ergs~s^{-1}}$) \citep{Rees84}. AGNs have various types depending on the selection method of the different wavelengths, such as, Seyfert galaxies \citep{Seyfert43}, Quasars \citep{Schmidt63}, and radio galaxies \citep{Fanaroff74}. The Seyfert galaxies are mainly identified by their optical emission lines. Seyfert 1 galaxies have broad Balmer emission lines whereas Seyfert 2 galaxies have only narrow Balmer emission lines \citep{Khachikian71,Khachikian74}. Also, there are some Seyfert galaxies showing features between Seyfert 1 and Seyfert 2 galaxies, e.g., intermediate Seyfert galaxies \citep{Osterbrock76}. These intermediate Seyfert galaxies are classified into Seyfert 1.2, Seyfert 1.5, Seyfert 1.8, and Seyfert 1.9 depending on the relative strength of H$\alpha$ and H$\beta$ emission lines; Seyfert 1.2/1.5 galaxies have strong broad Balmer component with a cusp of the narrow Balmer component \citep{Osterbrock77}. Seyfert 1.8 galaxies have weak broad Balmer component and strong narrow Balmer component while Seyfert 1.9 galaxies have only a weak broad H$\alpha$ component superimposed on a strong narrow H$\alpha$ component \citep{Osterbrock81}. However, \cite{Osterbrock83} discard Seyfert 1.2 and use Seyfert 1.5 for all the sources with a strong broad and narrow component. \cite{Whittle92} and \cite{
Winkler92} analyzed optical spectra of Seyfert galaxies using line ratio of H$\beta$ to [OIII] for the subclasses of the Seyfert 1 galaxies. However, the quantitative identification for Seyfert 1.9 is under investigation.

\cite{Osterbrock81} suggested that the observation characteristic of Seyfert 1.8/1.9 is due to dust reddening of Broad Line Regions (BLRs). However, NGC 2992 shows a variation of the weak broad H$\alpha$ component and the variation is not consistent with dust reddening; the variation is suggested to be caused by the intrinsic low-continuum state \citep{Trippe08}. \citet{Trippe10} showed that variation of 52\% (10/19) of Seyfert 1.8/1.9 is caused by low continuum state, while that of four sources might be caused by reddening of the broad line region.
In order to understand the possible mechanisms of Seyfert 1.9, collecting more observation sample is crucial. However, since we can only select Seyfert 1.9 galaxies by visual inspection due to its special characteristic. The identification process is time-consuming and has a potential bias for being inspected by different people.

With the advance of astronomical instruments, several large sky surveys, such as the Sloan Digital Sky Survey  \citep[SDSS;][]{york00}, Panoramic Survey Telescope and Rapid Response System  \citep[Pan-STARRS;][]{Chambers16}, and Palomar Transient Factory  \citep[PTF;][]{Law09,Rau09} have been conducted for more than one decade. These surveys bring the astronomy community to a big data era but also bring a new challenge in processing huge amount of observational data. The development of computer science provides possible solutions for big data, e.g., distinguishing star-forming galaxies and AGNs without the dominant characteristic of H$\alpha$ and [NII] emission lines by machine learning \citep{Teimoorinia18,  Zhang19}. Besides, identifying candidate AGNs by machine learning has been done by several authors \citep{Cavuoti14, S19, Faisst19}. \citet{Fraix2021} classify galaxy spectra by using an unsupervised method. Nowadays, one category of machine learning is deep learning. The deep learning has better ability to deal with huge data than machine learning. \citet{Leung19} shows to apply deep learning to stellar spectra for determining stellar abundance. Deep learning has several different algorithms for recognizing features. One popular algorithm is Convolution Neural Network (CNN) and is shown to success in image classification \citep{cire11,Krizhevsky17}. \citet{Krizhevsky17} is a successful case of image classification for classifying 1.2 million into 1000 classes and shows that by adding more convolution layers into model could reduce the classification error to 15\%. The feature of the CNN is multi layers and specific filter convoluted with input data; this is regarded as extracting features between different types of the input data. The learned information will be transmitted into next layer and the data transmission from low to high layer is forward pass. Another characteristic of the CNN is combination of back propagation and chain rule, the model will estimate output with respect to input data and the gradient will be sent from high layer back to low layer. This process is regarded as self-learning of the model. In astrophysics, there are some CNN examples;  \citet{Huertas15} classify galaxy morphology and predict features by a CNN model, and \citet{Pasquet2018} use a CNN model for classifying images of quasar light curves and find new quasar candidates.

In this paper, we build a 1-dimension (1D) CNN model to collect a large sample of the Seyfert 1.9 galaxies from Seyfert 2 galaxies.
The only difference between the optical spectra of Seyfert 1.9 and Seyfert 2 galaxies is the H$\alpha$ emission line. Seyfert 1.9 galaxies have a weak broad H$\alpha$ component superimposed by a strong narrow H$\alpha$ component whereas Seyfert 2 galaxies have only a strong narrow H$\alpha$ component. Taking benefit of our CNN model, we can collect more Seyfert 1.9 galaxies quickly than before by distinguishing spectral features of the Seyfert 1.9 and Seyfert 2 galaxies.
Besides, we provide line properties of the Seyfert 1.9 galaxies.
In section 2, we describe the algorithm of a CNN model. In section 3, we present data selection, our CNN model, and input of the CNN model. In section 4, we show training and testing results of our Seyfert sample. In section 5, we present the result of decomposition of the H$\alpha$ emission line of the Seyfert 1.9 galaxies. Finally, we discuss and summarize our results in Section 6 \&7. In this paper, we used $H_0=70$ km~s$^{-1}$~Mpc$^{-1}$,~$\Omega_m=0.3$,~$\Lambda_0=0.7$,~$q_0=-0.55$,~$k=0.0$.

\section{Convolution neural network} \label{sec:cnn}

CNN is an algorithm of machine learning and its characteristic is extracting features of input data through a specific sized filter convolution with input data. In the following, we give a brief introduction about neural network and how it works. We will also present our custom neural network model in the following.

\subsection{Neural Network}
In biological, a neuron means a nerve cell and can receive information from the environment and processes the information and send information to other neurons. In mathematics form, a neuron is defined as below:
\begin{equation}
y=f(\sum_ix_iw_i+b)
\end{equation}
where $x_i$ represents input value, $w_i$ represents weight, $b$ represents bias, $f$ is an activation function and $y$ represents the output from the neuron. The output of a neuron is applying an activation function to the summation of the dot result of the input value and weight and a bias value. An activation function is a nonlinear function and can be regarded as a learning behavior of the neuron. 

\subsection{Basic layer in a Convolution Neural Network}

A typical CNN consists of an input layer, several hidden layers which are convolution layers, pooling layers ,and a or more fully-connected layers in the ends before the output layer. A convolution layer means that a filter with a specific size and a set of weight values will across the input data to compute the dot result of the filter and input data. After the convolution of the filter and data, it can highlight the feature of input data.  Therefore, a convolution layer can detect the possible features of objects from the dotting result of input data and filter. The pooling layer has a specific filter and the filter will read input data with the same length. Every reading will output a result and the result depends on the pooling method. For example, max-pooling is to output the maximum value in every reading of the filter. The function of the pooling layer is to retain the dominant feature of the input data and discard the other information; this can reduce the data length and parameter of the model to make computation fast without losing primary information. The fully-connected layer is usually placed before the output layer. The function of the fully-connected layer is to flat all input neurons and every neuron will be calculated with weight values. The calculation in this layer makes sure all information will be transmitted. The neurons in different layers can connect but the neurons do not connect to the neurons in the same layer. This means that every feature learned from the previous layer will be combined in the fully-connected layer. Then, according to comprehensive information from all layers, the fully-connected layer will output the probability of different object classes in the end.

\subsection{Parameters of a convolution neural network model}

The CNN model has two passways, one is the forward pass and another is the backward pass called backpropagation. The forward pass is to estimate the output result from input data and the transfer direction is from the low layer to the high layer. Backpropagation combines gradient descent and chain rule and to transmit the gradient, which is the output respect to the input value, from high layer to low layer.

In the convolution neural network model, weight and bias are the parameters of the model. To evaluate how different between the model with update parameters and the target is loss function. A small loss represents that the difference between predicted output and target output is small, indicating that the output results are closed to target and the classification result of the model is good. A large loss represents that the difference between model and output is large, indicating the fitting result is bad and the model can not classify objects correctly. In the CNN model, it uses the gradient descent method to find the minimum difference between the model and target and optimizing loss function. Estimating the gradient of the loss to parameter provides trend direction in parameter space and helps the model obtain the appreciate parameter.

We present how gradient descent is applied in the CNN model below.
\begin{equation}
w'=w-\eta\frac{\partial L}{\partial w}
\end{equation}
where $w'$ is updated weight from $w$, $\eta$ is the learning rate, $L$ is loss.  In this work, we use the Cross-Entropy Loss because this loss is appropriate for binary or multiple classifications. The optimizer is the process to find the best parameter of the model that has a minimum loss. In this work, we use the Stochastic Gradient Descent (SGD). This will randomly select a sample from the whole as the beginning for calculating the gradient. It will update the gradient by a small sample selected randomly from whole data in every iteration. The learning rate decides the quantity of the update step, a small step will update slowly but steadily. A big step will cause unstable during the update process. The epoch in a training process represents iteration times for the training model.

\begin{table}
\caption{1D CNN architecture}
\label{tab:cnn}
\centering
\begin{tabular}{lcccccc}
\toprule
Layer & Type & Channel$_\mathrm{input}$ & Channel$_\mathrm{output}$ & Kernel size & Stride & Activation \\
\hline
1 & Convolution & 1 & 10 & 200 &1 &ReLU\\
2 & Pooling &  & & 2 & 2&\\
3 & Convolution & 10 & 30 & 100 &1&ReLU\\
4 & Pooling &  & & 5 &2&\\
5 & Convolution & 30 & 36 & 56& 1 &ReLU\\
6 & Linear & 3384 & 94 & & &ReLU\\
out & Linear & 94 & 2 &\\
\toprule
\end{tabular}
\end{table}

\section{Data selection and analysis} \label{sec:data_sel}

\subsection{Data selection}

We select our Seyfert sources from \cite{Chen19}. This sample includes 54694 Seyfert 2 galaxies and 745 Seyfert 1.9 galaxies with 0 $<$ z $<$ 0.2.
Our samples criteria are
S/N of H$\alpha > 3$, S/N of [OIII] $>$ 5 , and log $L_\mathrm{{[OIII]}}$~[ergs/s] $> 40.125$.  We obtain the optical spectra from the Sloan Digital Sky Survey Data Release 10 \citep[SDSS DR10;][]{Ahn14}. In our Seyfert 1.9 sample, some of them might include Seyfert 1.2, Seyfert 1.5, and Seyfert 1.8. We inspect  H$\beta$ emission line of the 745 sources and find 641 pure Seyfert 1.9 galaxies. Therefore, we have two training sets. Training set 1 uses 745 intermediate Seyfert and 56494 Seyfert 2 sources. Training set 2 uses 641 Seyfert 1.9 and 56494 Seyfert 2 sources. In order to learn nature difference between Seyfert 1.9 and Seyfert 2 galaxies instead of their distribution, we choose the same number for Seyfert 1.9 and Seyfert 2 as our training data set. Therefore, we use 300 intermediate Seyfert and/or Seyfert 1.9 and 300 Seyfert 2 samples in the training process. For the rest sources, we have 445 intermediate Seyfert and/or 341 Seyfert 1.9 and 53494 Seyfert 2 galaxies as the test data set. We show the detail amount of training and test number in different training sets in Table.~\ref{tab:tt}

We only use a segment of a spectrum as our input because this range covers the H$\alpha$ emission feature in the optical spectrum. The segment range is from 6400\AA~to 6700\AA~and all spectra are shifted back to the rest frame.
In order to focus on the shape of the emission line, we normalize every spectrum to its peak value to make the value range between 0 and 1. This normalization can make our model more easy to read without confusion.

\subsection{Neural Network architecture}

Our customized model is a 1-dimension CNN, which means the input format is a sequence array. We only input flux values of spectra with wavelength. We show the architecture in Table \ref{tab:cnn}. This model has 6 layers, which include 1 input layer, four hidden layers ,and two linear layers. The final output is 2 channels and this represents that the output results are either Seyfert 1.9 or Seyfert 2 galaxies. For the training model, we set batch size = 30. The batch size relates to SGD and represents how many sources will be sent into the model every time. We set a constant learning rate=0.01 during learning and the total epoch is 100.

\section{Results} \label{sec:results}

\begin{figure}[ht]
\begin{center}
\includegraphics[scale=0.54,angle=0]{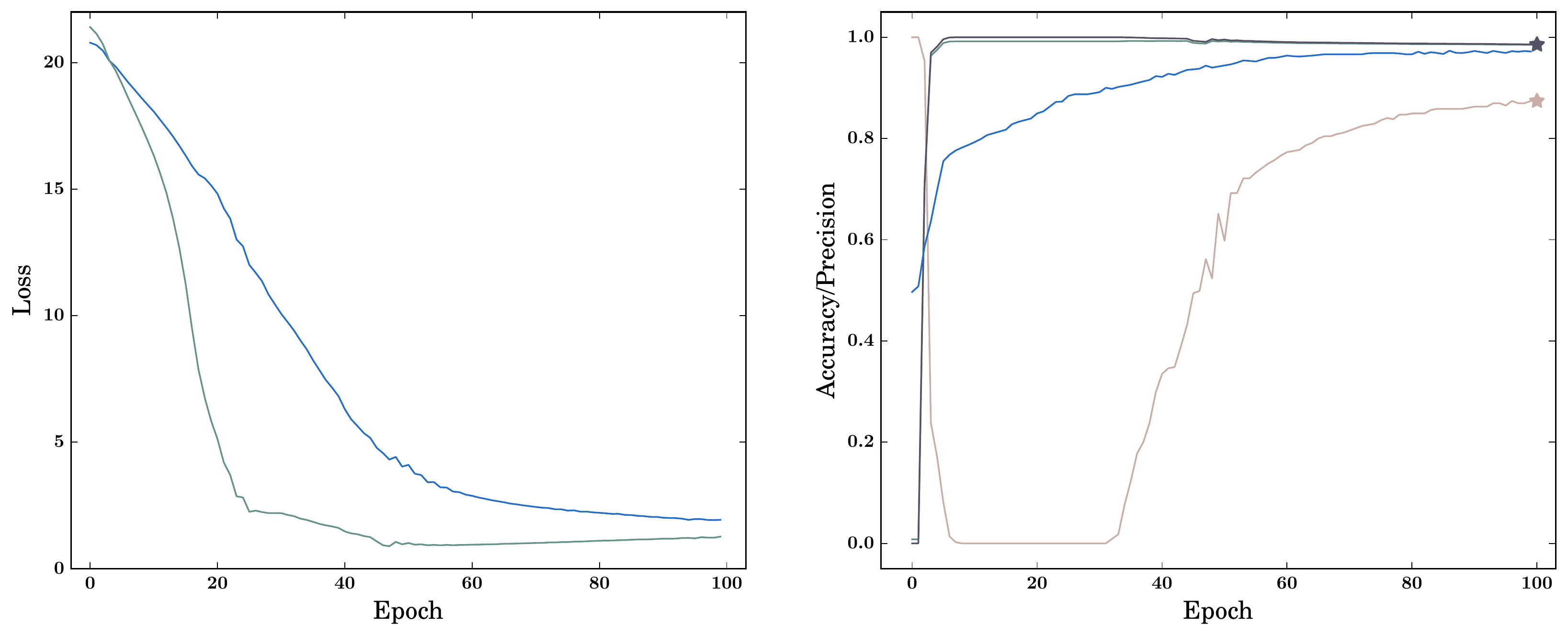}
\caption{Loss and accuracy/precision as a function of the epoch of training set 1. Left: loss value as a function of epoch. Right:
accuracy/precision value as a function of epoch. 
Blue represents the training process and the green represents the valid process. Light pink represents the precision of intermediate Seyfert and dark purple represents that of Seyfert 2. Star represents the test value.}
\label{fig:1}
\end{center}
\end{figure}

A loss value represents the difference between a predicted output from the model and a target. In our case, we have 300 training sources for two Seyfert types, respectively, with a batch size =30. In each epoch, the data will be sent into the model in 10 times and every time has 30 loss values. We sum 30 loss values in every batch and average total 300 loss values as a represent value for one epoch. The variation loss value with epochs is usually called learning curve to indicate how well the model learns. We show the learning curves of training set 1 in Fig.~\ref{fig:1}. The valid sample is used to evaluate a given model and to fine-tune the hyperparameters. Training model would not learn from the valid sample. We find the loss values of training and valid sample decrease with epoch, representing the differences between target and predicted result decrease with iteration. After 50th epoch, the learning curves of both training and valid converge to a stable point, indicating our model reaches its optimal fit. This result indicates that our model learns well and finds the best parameter in parameter space during learning process. The final loss values of training and test are 1.93 and 1.27, respectively. In order to know the classification ability of our training model, we estimate the accuracy value of training and test sample. In every epoch, all sources will be classified by our model and the accuracy is estimated from the classified correct sources divided by total sources in every epoch.  We show the accuracy/precision as a function of epoch in Fig.~\ref{fig:1} and test results in Table~\ref{tab:tt}. We find the accuracy of training increases with epoch and the accuracy of valid fluctuates around 0.99 after epoch=3. The test accuracy after 100 epochs is 98\%. For more detail, we would like to know how many sources are classified as their label types and estimate the precision of the two Seyfert types. For Seyfert 1.9, the precision reaches 87\% after 100 epochs. This represents that our model classifies correctly 389 Seyfert 1.9 samples out of 445 Seyfert 1.9 samples after iterating 100 epochs. For Seyfert 2, the precision is 98\% after iterating 100 epochs and that is our model classifies correctly 53628 Seyfert 2 samples out of 54394 Seyfert 2 samples.

\begin{table}
\caption{Summary of the training and test results}
\label{tab:tt}
\raggedleft
\resizebox{\textwidth}{!}{
\begin{tabular}{rcccccccc}
\toprule
 Training set& Seyfert type & Train sample & Test sample & precision  & Accuracy  & 2nd Test sample & precision & Accuracy \\
\hline
\multirow{2}{1.5cm}{1} & intermediate Sey & 300 & 445 & 87\% & \multirow{2}{0.6cm}{98\%} & 641* & 86\% &\multirow{2}{0.6cm}{98\%} \\&Sey2 & 300 & 54394 & 98\% & & 54694 & 99\% \\ [0.2cm]
\multirow{2}{1.5cm}{2} & Sey1.9 & 300* & 341* & 84\% & \multirow{2}{0.6cm}{98\%} &  &  \\&Sey2 & 300 & 54394 & 98\% & &  \\ [0.2cm]
\multirow{2}{1.5cm}{3} & Sey1.9 & 600* & 697* & 91\% & \multirow{2}{0.6cm}{99\%} &  &  \\&Sey2 & 600 & 53328 & 99\% & &  \\
\toprule
\multicolumn{9}{l}{%
  \begin{minipage}{13.5cm}%
    \tiny Note: * means the pure Seyfert 1.9 galaxies. \\
    That is to say these spectra do not have broad H$\beta$ emission line and have relatively weak broad H$\alpha$ componet.%
  \end{minipage}%
}\\
\end{tabular}}
\end{table}

\begin{figure}[ht]
\begin{center}
\includegraphics[scale=0.54,angle=0]{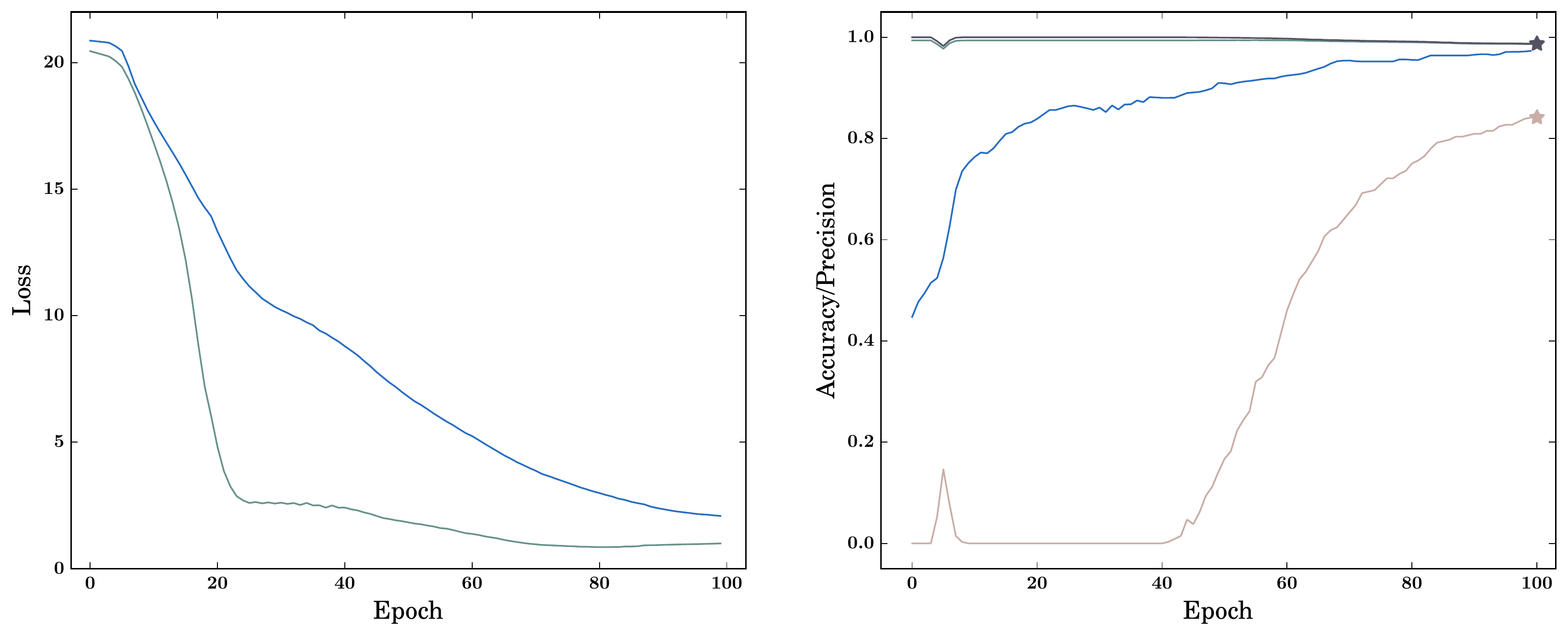}
\caption{Loss and accuracy/precision as a function of epoch of training set 2. Left:
loss value as a function of the epoch. Right: accuracy value as a function of epoch. Blue represents the training process and the green represents the valid process. Light pink represents the precision of Seyfert 1.9 and dark purple represents that of Seyfert 2. Star represents the test value.}
\label{fig:dataset2}
\end{center}
\end{figure}

In training set 2, we use 641 pure Seyfert 1.9 sample as our Seyfert 1.9 sample to train our model and test if the model still can discern the characteristic of weak broad component of Seyfert 1.9 galaxies among the Seyfert 2 galaxies. We use 300 Seyfert 1.9 and Seyfert 2, respectively, as our training sample. The remaining Seyfert 1.9 sources reduce to 341 as the test data source. We use the same initial parameters,  batch size=30 and learning rate=0.01, for training our model again. We show the results of the training set 2 in Fig.~\ref{fig:dataset2} and Table~\ref{tab:tt}. The loss values of both training and valid are decreased with epoch. The final training loss is 2.07 and the final valid loss is 1. The test accuracy for training set 2 is 98\%. The precision of the Seyfert 2 is 98\% after 100 epoch. However, the precision of the Seyfert 1.9 after 100 epoch is 84\% and that is to say this model can classify 287 Seyfert 1.9 correctly from 341 Seyfert 1.9 galaxies. This indicates that our training set 2 model can still discern pure Seyfert 1.9 galaxies from Seyfert 2 galaxies.

In this work, we make a CNN model discern the spectrum with a component (Seyfert 2) or with two components (intermediate Seyferts) on H$\alpha$ emission line. Although all intermediate Seyferts have two components on their H$\alpha$ emission line, the ratio between broad and narrow component is different. \citet{Stern12} indicate that peak flux density ratio of broad H$\alpha$ to narrow H$\alpha$ for Seyfert 1.5 is about 0.5 and for Seyfert 1.9 is about 0.05. This indicates that a Seyfert 1.9 has a weaker broad H$\alpha$ component than the other intermediate Seyferts. In training set 1, we use all intermediate Seyferts for distinguishing two-component features on H$\alpha$ emission. In training set 2, we only use Seyfert 1.9 galaxies as training sample to investigate if the trained model can discern weak broad H$\alpha$ component. We find the models trained by intermediate Seyferts with different strength of the broad H$\alpha$ component and Seyfert 1.9 have similar precision results.
We want to know if the model trained by training set 1 can discern the Seyfert 1.9 which has relative weak broad component than other intermediate Seyfert galaxies and check if this feature would increase  classified difficulty for the training set 1 model. We apply the training set 1 model to test the second test sample. 
 This time, we have 641 Seyfert 1.9 galaxies and 54694 Seyfert 2 galaxies as our second test sample and show the test results in Table~\ref{tab:tt}. The accuracy for this test is 98\% and that is the model of training set 1 can recognize 54481 sources correctly from 55335 sources. For Seyfert 1.9, the precision is 86\% and this means the model of training set 1 can classify correctly 553 Seyfert 1.9 sources from 641 Seyfert 1.9 sources. For Seyfert 2, training set 1 model can discern 53928 sources correctly from 54694 sources. This result indicates that although our training set 1 model is trained by stronger broad H$\alpha$ sample, it still can recognize   Seyfert 1.9 which has weak broad H$\alpha$ component among the Seyfert 2 galaxies with a precision of 86\%. We notice 255 out of 641 Seyfert 1.9 sources are repeated sources in training sample of the training set 1. In addition, we notice although the training set 1 model can classify most Seyfert 2 sources correctly in the second test sample, it remains 766 Seyfert 2 galaxies classified as Seyfert 1.9 galaxies in this test. Thus, we inspect the 766 classified wrong Seyfert 2 sources and find that 10 of 766 are damage spectra, 707 spectra have weak broad H$\alpha$ component, and 44 spectra have no H$\alpha$ emission or no broad component. We also inspect visually the H$\beta$ emission line of 707 spectra and find 656 sources are Seyfert 1.9 galaxies and 53 sources are other intermediate Seyfert galaxies. These results indicate that our model can pick out the Seyfert 1.9 galaxies which were missed by visual inspection from Seyfert 2 galaxies and obtaining Seyfert 1.9 galaxies among Seyfert 2 galaxies via deep learning is practicable.

\begin{figure}[ht]
\begin{center}
\includegraphics[scale=0.54,angle=0]{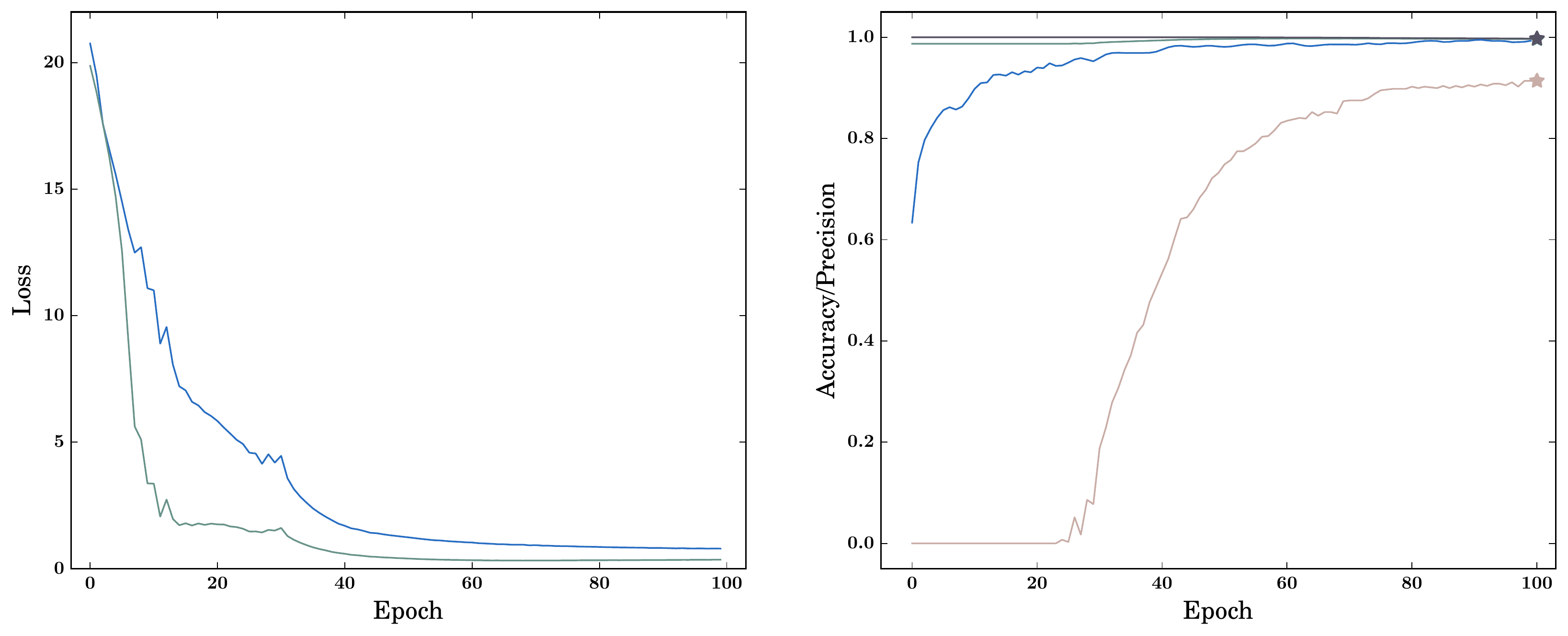}
\caption{Loss and accuracy/precision as a function of the epoch of training set 3. Left:
loss value as a function of epoch. Right: accuracy/precision value as a function of epoch. The color symbols are the same in Fig.~\ref{fig:dataset2}.}
\label{fig:dataset3}
\end{center}
\end{figure}

From previous result, our model picks out additional 655 Seyfert 1.9 sources. Therefore, we further set a training set 3 by using more pure Seyfert 1.9 sources as training sources to train our model again. The total Seyfert 1.9 sample is 1297 (641 + 656) sources. We remove 766 sources which are either damage spectra or Seyfert 1.9 spectra from 54694 Seyfert 2 sample. Thus, we have 1297 Seyfert 1.9 and 53928 Seyfert 2 galaxies in training set 3. This time, we double our training sources to 600 Seyfert 1.9 and 600 Seyfert 2 galaxies and the test source is 697 Seyfert 1.9 and 53628 Seyfert 2 galaxies. We summary sources amount of training set 3 in Table.~\ref{tab:tt} and show the training results in Fig.~\ref{fig:dataset3} and Table.~\ref{tab:tt}. The final training loss is 0.79 and final valid loss is 0.36. We find that the loss values of training and valid decrease with epoch and converge to a constant value after epoch = 60. This plateau indicates our model reaches its global minimum point and the fitting result is good. The test accuracy is 99\% and it means our model model can classify 53827 sources correctly among 54025 sources. For classifying Seyfert 1.9 galaxies, our model has a precision of 91\% and it indicates our model can classify 637 Seyfert 1.9 sources correctly among 697 Seyfert 1.9 sources. For Seyfert 2, our model has a precision of 99\% for classifying 53190 Seyfert 2 sources correctly among 53328 Seyfert 2 sources.

\section{Emission line properties of the Seyfert 1.9 galaxies} \label{sec:emline}

\begin{figure}[ht]
\begin{center}
\includegraphics[scale=0.7,angle=0]{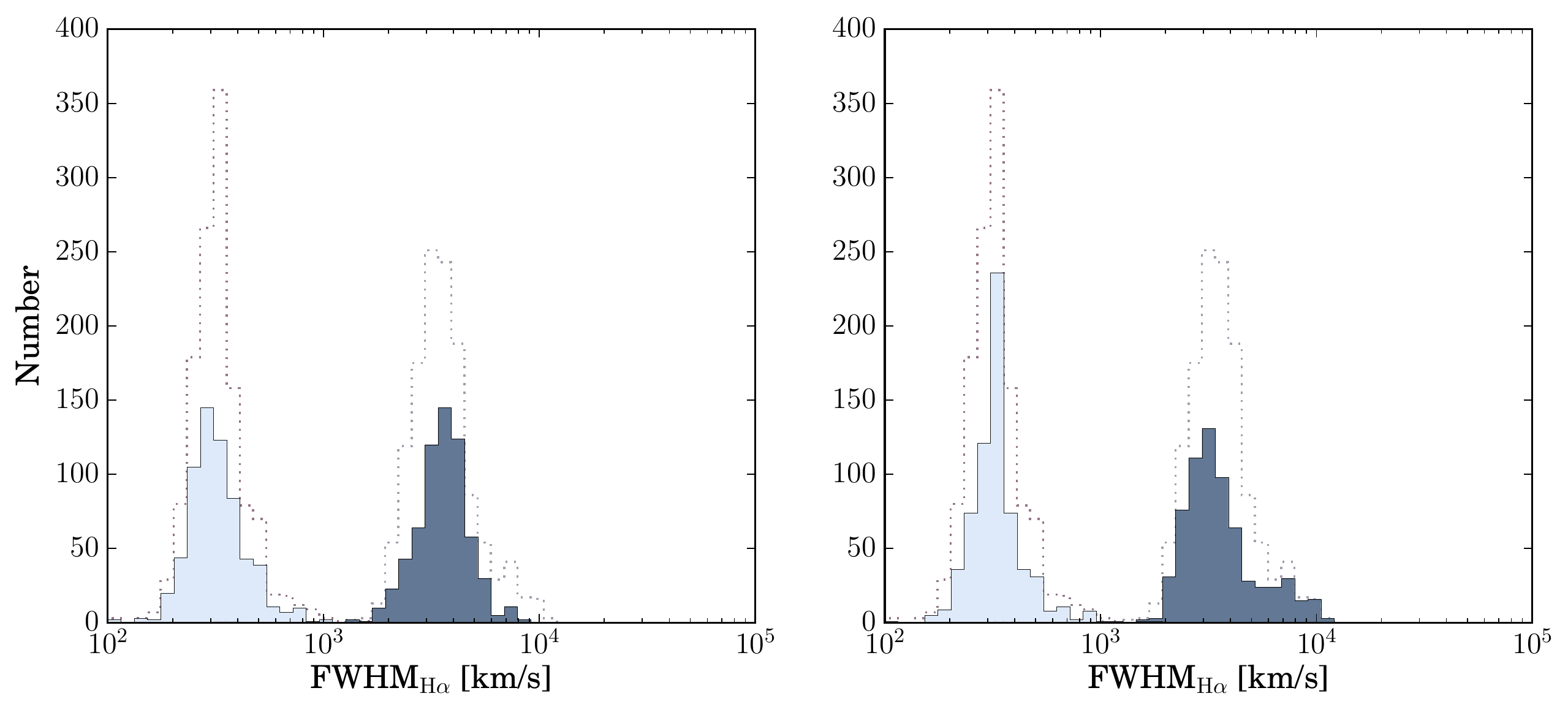}
\caption{FWHM distributions of our Seyfert 1.9 sources. Left: human-selected Sy1.9. Right: machine-selected Sy1.9. Dot-line represents the total 1297 Seyfert 1.9 sources. Light blue represents the narrow H$\alpha$ component. Dark blue represents the broad H$\alpha$ component.}
\label{fig:fwhm_dis}
\end{center}
\end{figure}

We decompose the H$\alpha$ emission line of 1297 Seyfert 1.9 galaxies by fitting two Gaussian components as a broad and narrow component. More detail of the fitting is presented in Appendix \ref{appendix:a}. We derive line width of broad and narrow component from fitting result and show full width at half maximum (FWHM) distribution in Fig.~\ref{fig:fwhm_dis}. We find the narrow H$\alpha$ component of the 1297 Seyfert 1.9 galaxies distributes from 100 to 1000 [km/s] whereas the broad H$\alpha$ component distributes from 1000 to 10000 [km/s]. The distribution of the broad H$\alpha$ component has averagely one order larger than the narrow H$\alpha$ component. For more details between two Seyfert 1.9 samples, we use ``human-selected Sy1.9'' for the 641 Seyfert 1.9 sources that are selected by human inspection and ``machine-selected Sy1.9'' for the 656 Seyfert 1.9 sources that are selected by our CNN model. In the broad H$\alpha$ component distributions, we find the peak of the  human-selected Sy1.9 sample is slightly higher than that of the machine-selected Sy1.9 sample and we notice the machine-selected Sy1.9 sample has an extending tail toward the higher velocity end.

\begin{figure}[ht]
\begin{center}
\includegraphics[scale=0.7,angle=0]{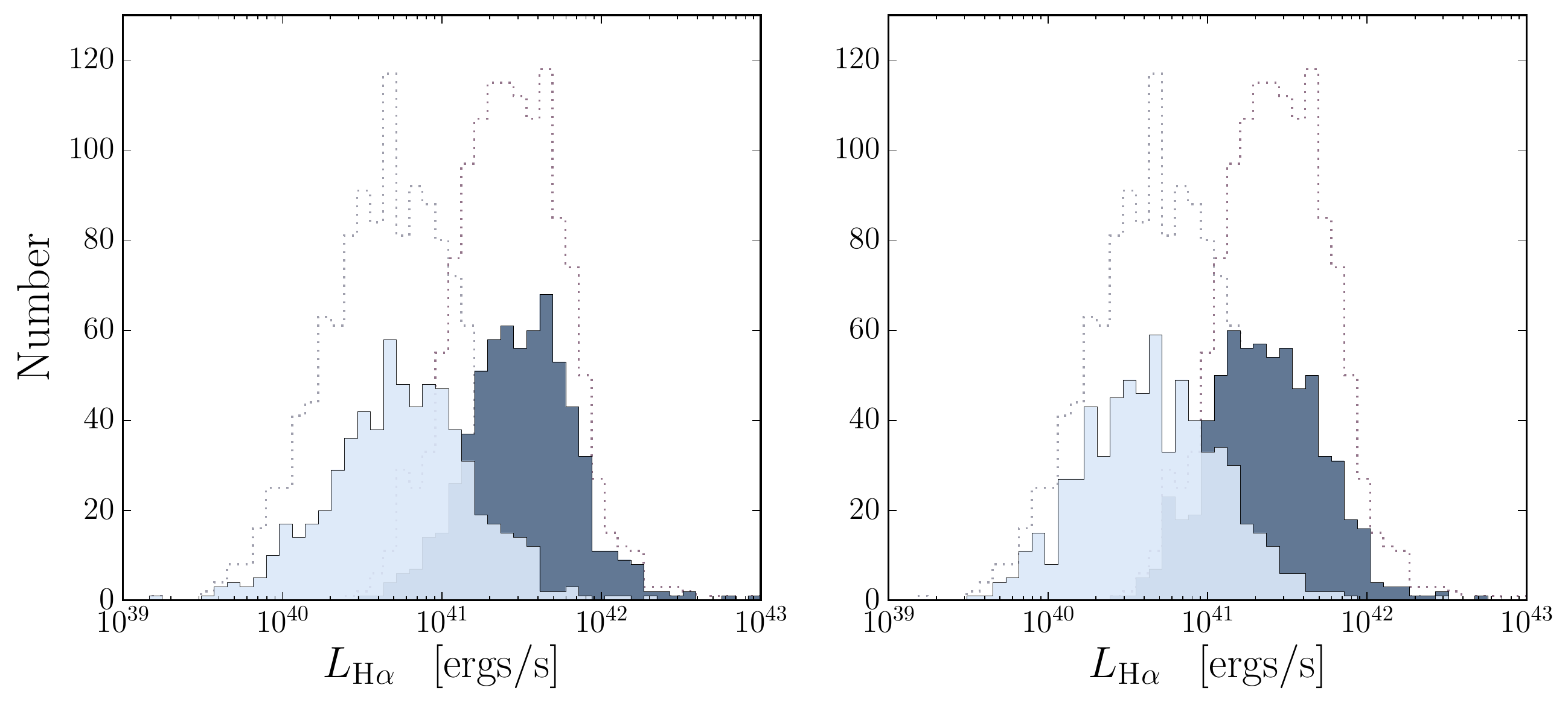}
\caption{H$\alpha$ luminosity distributions of our Seyfert 1.9 sources. Left: human-selected Sy1.9 sample. Right: machine-selected Sy1.9 sample. Dot-line represents the total 1297 Seyfert 1.9 sources. Light blue represents the narrow H$\alpha$ component. Dark blue represents the broad H$\alpha$ component.}
\label{fig:Lha_dis}
\end{center}
\end{figure}

We also estimate H$\alpha$ luminosity of broad and narrow component and show the results in Fig.~\ref{fig:Lha_dis}. The luminosity distribution of narrow H$\alpha$ component of both Seyfert 1.9 samples span from 10$^{39}$ to 10$^{42}$ [ergs/s] with a mean value $\approx$ 10$^{40}$ [ergs/s] while the luminosity distribution of broad H$\alpha$ component spans from 10$^{40}$ to 10$^{43}$ [ergs/s] with a mean value $\approx$ 10$^{41}$ [ergs/s]. We find that the luminosity distribution of broad H$\alpha$ component has one order larger than that of narrow H$\alpha$ component in our Seyfert 1.9 sample. We show the mean values of FWHM$_\mathrm{H\alpha}$ and H$\alpha$ luminosity in Table.~\ref{tab:Sey_em}. We find that the machine-selected Sy1.9 sample have slight higher mean value of the FWHM$_\mathrm{H\alpha}$ than the human-selected Sy1.9 sample. However, the mean H$\alpha$ luminosity of the machine-selected Sy1.9 sample is relatively lower than that of the human-selected Sy1.9 sample. This result suggests that a source with a broader component is not necessary to have a higher luminosity.

\begin{table}
\caption{Comparison of the Seyfert 1.9 sample}
\label{tab:Sey_em}
\centering
\begin{tabular}{lccc}
\toprule
  & human-selected Sy1.9 sample & machine-selected Sy1.9 sample \\
\hline
 Number & 641  & 656 &  \\
FWHM$_{\mathrm{H\alpha,broad}}$  & 3682.09  & 3882.76 & [km/s] \\
FWHM$_{\mathrm{H\alpha,narrow}}$  & 336.09 &  345.39 & [km/s]  \\
L$_{\mathrm{H\alpha,broad}}$  & 4.45$\times 10^{41}$ & 3.25$\times 10 ^{41}$ & [ergs/s] \\
L$_{\mathrm{H\alpha,narrow}}$  & 9.47$\times 10 ^{40}$ & 7.75$\times 10 ^{40}$ & [ergs/s] \\
\toprule
\end{tabular}
\end{table}

\begin{figure}
\centering
\subfloat{
\includegraphics[width=\linewidth]{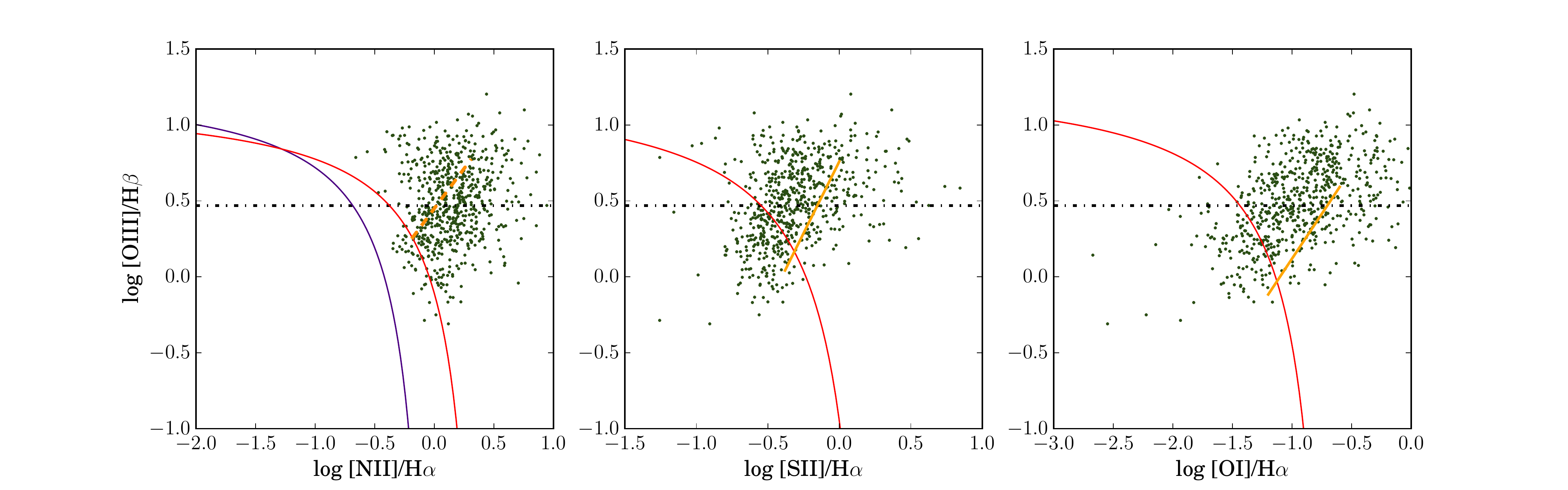}}

\subfloat{\includegraphics[width=\linewidth]{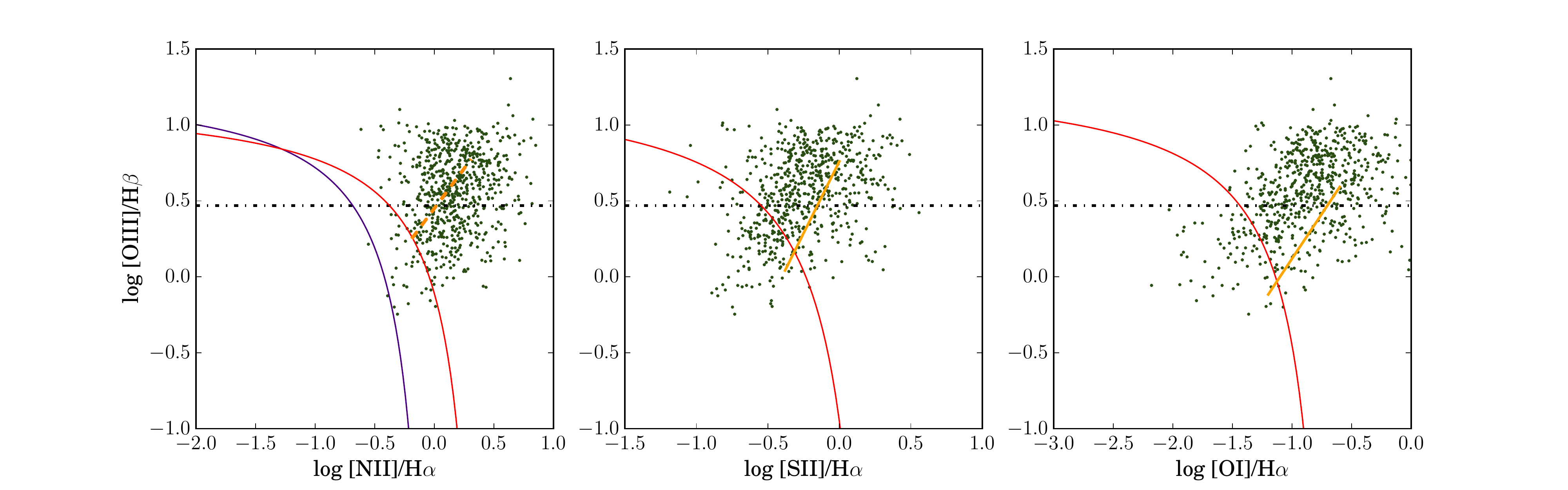}}

\caption{BPT diagram for our Seyfert 1.9 sample. Top: human-selected Sy1.9 sample. Bottom: machine-selected Sy1.9 sample. The red line represents the definition of the starburst limit \citep{Kewley01}. The dashed-orange line in [NII]/H$\alpha$ diagram represents  an experience division for Seyfert-LINER \citep{Schawinski07}. The orange line represents the Seyfert-LINER line \citep{Kewley06}. The purple solid line represents the AGN definition \citep{Kauffmann03}. The black dashed line represents the ratio of [OIII]/H$\beta$= 3.}
\label{fig:BPT}
\end{figure}

The ``Baldwin, Phillips \& Terlevich'' (BPT) diagram is used to distinguish different ionization mechanism of nebular gas \citep{Baldwin81}. Since the BPT diagram is for the narrow emission line, we only use the flux of narrow H$\alpha$ component to plot the BPT diagram. We show the results in Fig.~\ref{fig:BPT} and find that both of the two Seyfert 1.9 samples have most sources inside the AGNs region in the [NII]/H$\alpha$, [SII]/H$\alpha$, and [OI]/H$\alpha$ diagram. In human-selected Sy1.9 sample, there are 4.9\% (32/641) sources in the [NII]/H$\alpha$ diagram, 29.1\% (187/641) sources in the [SII]/H$\alpha$ diagram, and 16.6\% (107/641) sources in the [OI]/H$\alpha$ diagram outside the AGN region. In the machine-selected Sy1.9 sample, there are 4.7\% (31/656) sources in the [NII]/H$\alpha$ diagram, 19.9\% (131/656) sources in the [SII]/H$\alpha$ diagram, and 10.2\% (67/656) sources in the [OI]/H$\alpha$ diagram outside the AGN region. These results indicate that our Seyfert 1.9 sources are dominated by AGNs and are ionized by AGNs instead of stellar. We also compare the BPT diagram of our Seyfert 1.9 with that of Seyfert 2 galaxies in \citet{Chen19}. We find the distributions of [SII]/H$\alpha$ and [OI]/H$\alpha$ diagrams are similar to  Seyfert 2 galaxies. This result indicates that Seyfert 1.9 and Seyfert 2 galaxies are ionized by similar narrow line region. However, we notice in the [NII]/H$\alpha$ diagram the distributions of Seyfert 1.9 and Seyfert 2 are slightly different. We find our Seyfert 1.9 galaxies have relatively low [OIII]/H$\beta$ and high [NII]/H$\alpha$ than the Seyfert 2 galaxies. The slightly strong [NII]/H$\alpha$ in Seyfert 1.9 might be related to the stellar population in the host galaxy \citep{yu13}.

\section{Discussion} \label{sec:discu}

\begin{figure}[ht]
\begin{center}
\includegraphics[scale=0.65,angle=0]{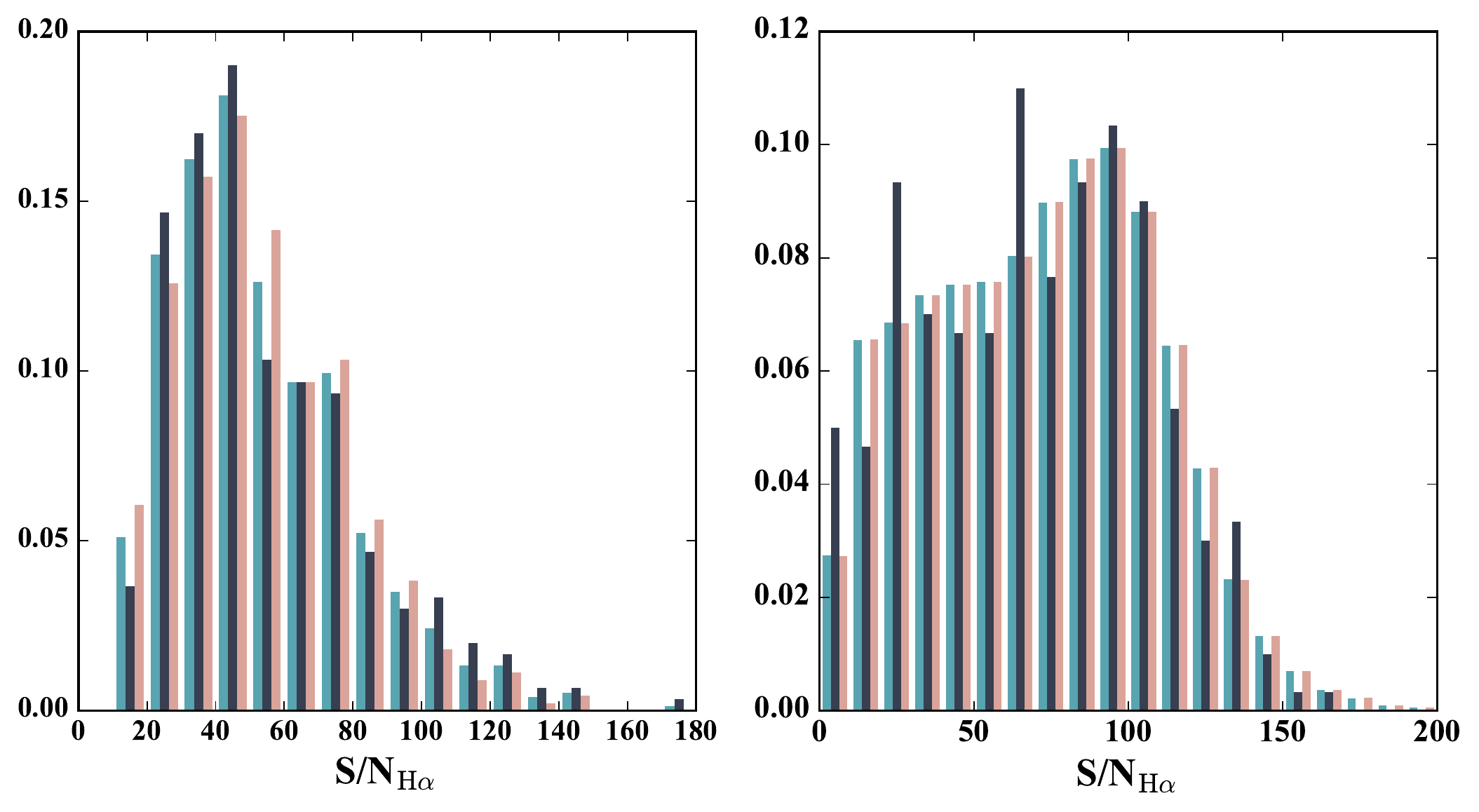}
\caption{S/N of H$\alpha$ emission line of our Seyfert sources in training set 1. Left: intermediate
Seyfert galaxies. Right: Seyfert 2 galaxies. Teal represents all sources. Dark gray represents training sources. Light pink represents test sources.}
\label{fig:sn1}
\end{center}
\end{figure}

\begin{figure}[ht]
\begin{center}
\includegraphics[scale=0.65,angle=0]{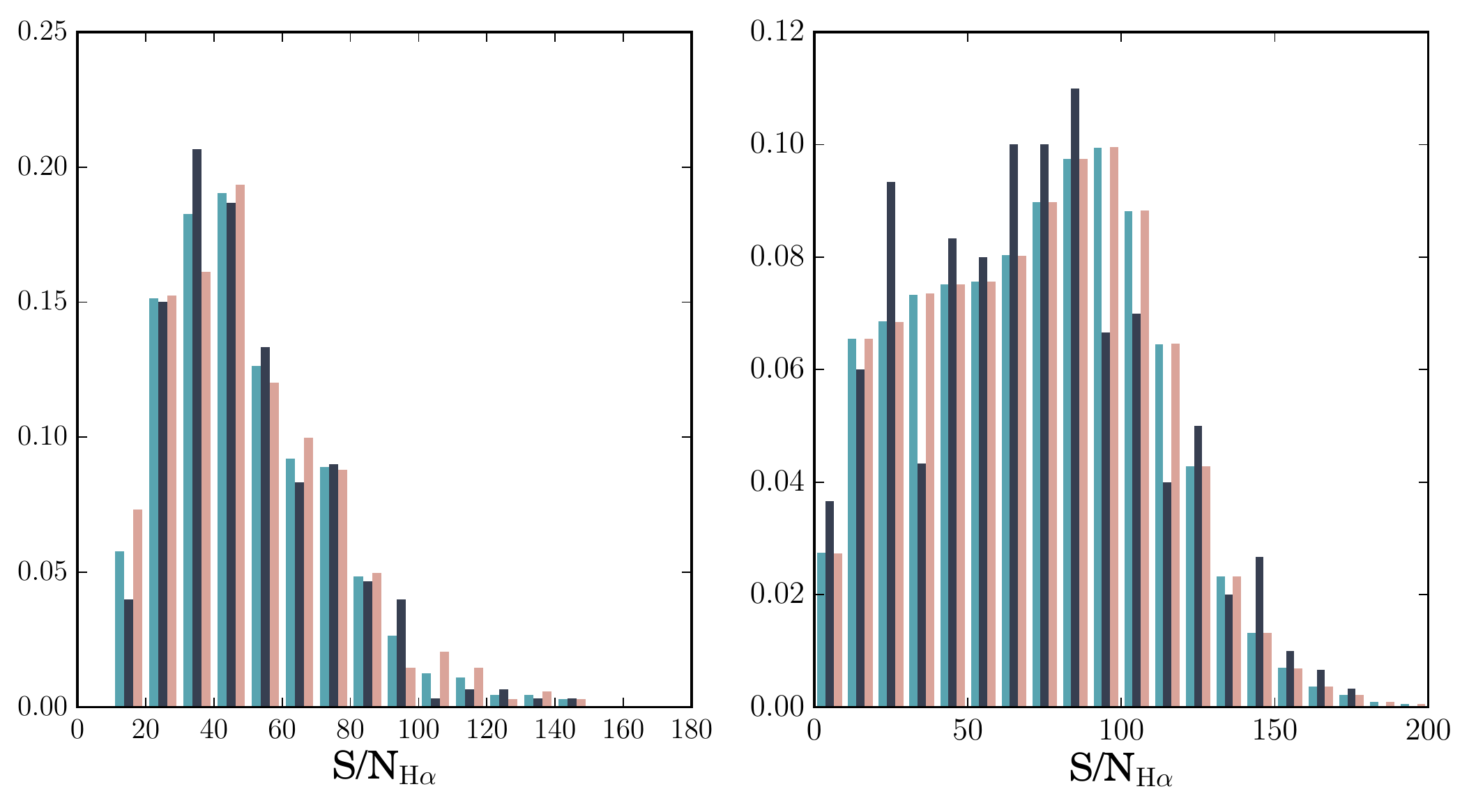}
\caption{S/N of H$\alpha$ emission line of our Seyfert sources in training set 2. Left:
Seyfert 1.9 galaxies. Right: Seyfert 2 galaxies. Teal represents all sources. Dark gray represents training sources. Light pink represents test sources.}
\label{fig:sn2}
\end{center}
\end{figure}

\begin{figure}[ht]
\begin{center}
\includegraphics[scale=0.65,angle=0]{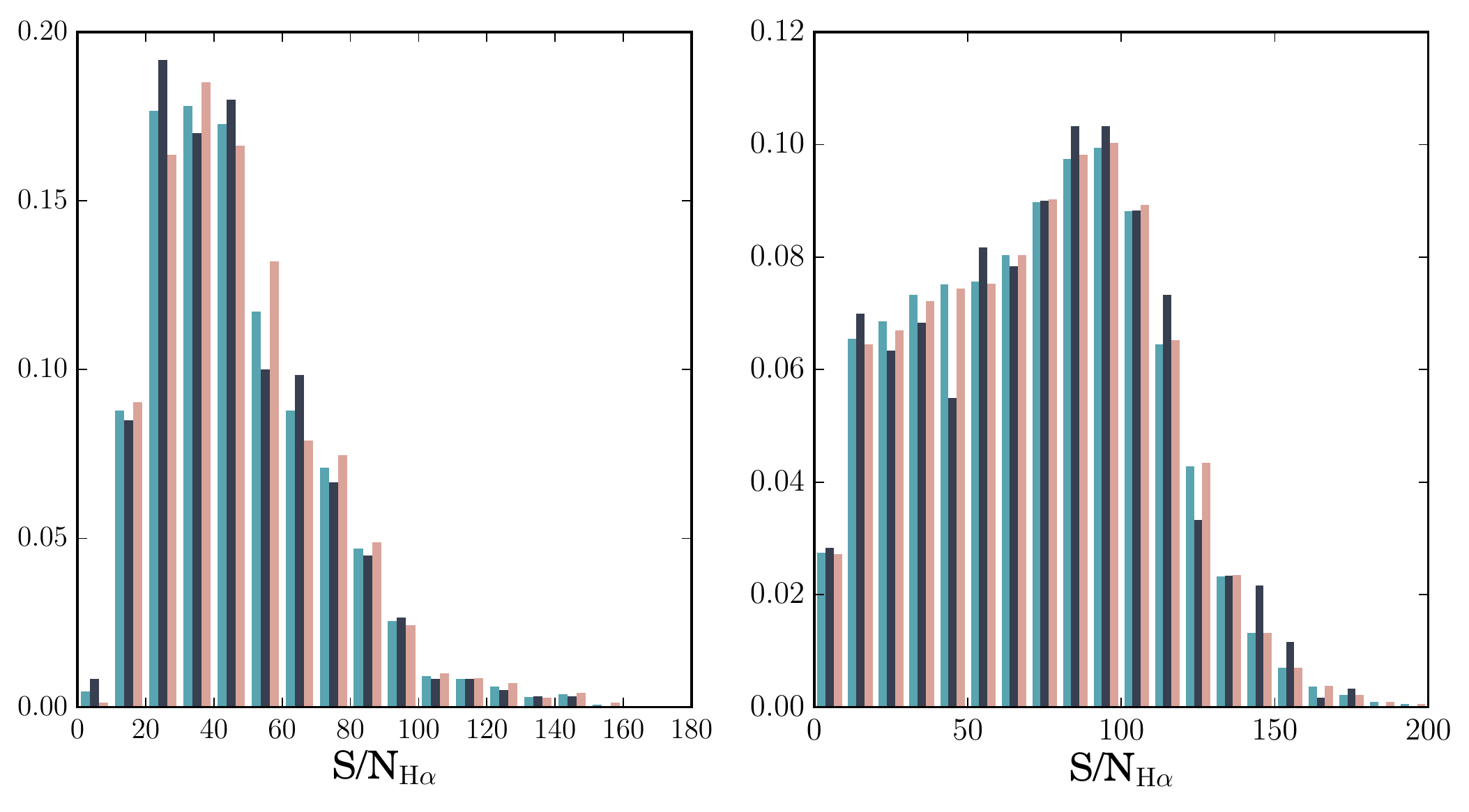}
\caption{S/N of H$\alpha$ emission line of our Seyfert sources in training set 3. Left:
Seyfert 1.9 galaxies. Right: Seyfert 2 galaxies. Teal represents all sources. Dark gray represents training sources. Light pink represents test sources.}
\label{fig:sn3}
\end{center}
\end{figure}

We would like to know if the Seyfert 1.9 galaxies that are classified by our model as Seyfert 2 galaxies are related to low-quality spectra S/N and we check the S/N distribution of training and test data. We show the S/N distribution of the training set 1 in Fig.~\ref{fig:sn1}. We find that the S/N distribution of the whole population is not different from that of training data and test data for the Seyfert 1.9 and Seyfert 2 sample. Besides, we show the S/N distributions of the training set 2 in Fig.~\ref{fig:sn2}. We find that the S/N distribution of the whole population of the Seyfert 1.9 does not show a significant difference from that of training and test data. These results indicate that the sources that are classified wrong by our model are not related to the S/N and suggest that our model can recognize Seyfert 1.9 spectrum and Seyfert 2 spectrum with low S/N. The possible reasons for classifying wrong are either insufficient training sources or needing more layers in our model to recognize the fine distinction between the spectra of Seyfert 1.9 and Seyfert 2 galaxies. Due to the {\bf 2$^\mathrm{nd}$} test result from training set 1, we have more Seyfert 1.9 sources. We use more Seyfert 1.9 sources in our training set 3 and we find adding more training sources can improve the performance of our model with decreased loss and an increased precision of Seyfert 1.9. We show the S/N distributions of our sources of training set 3 in Fig.~\ref{fig:sn3}. We find the S/N distributions of our training and test sources have a similar distribution. These results indicate that S/N distributions are not related to our test results and our model can deal with low S/N sources again.

V\'{e}ron catalog collects intermediate Seyferts from literature and has 151 Seyfert 1.9 galaxies with redshift range from 0 to 0.2 \citep{Veron10}. We find our Seyfert 1.9 sample is almost ten times as the V\'{e}ron 13th catalog. We also provide H$\alpha$ fitting results of the Seyfert 1.9 galaxies. \citet{Ho97} fit 486 candidate spectra with a different fraction of the broad component for determining the existence of a broad H$\alpha$ and they found $\approx$ 16\% of sources have a broad H$\alpha$ component. Their sources have L$_{\mathrm{H\alpha, broad}}~\approx$ ~10$^{39}$ [ergs/s] and FWHM$_{\mathrm{H\alpha, broad}}~\approx$~2200 [km/s]. \citet{Stern12} fit the candidate spectra and find the excess flux near H$\alpha$ emission line. They have 3579 sources with the L$_{\mathrm{H\alpha, broad}}$ ranging from 10$^{40}$ to 10$^{44}$ [ergs/s]. We find both methods described above have to fit amount candidate spectra at first and are usually time-consuming. The classification of sources might depend on the fitting result. By comparing the luminosity of broad H$\alpha$, we find our Seyfert 1.9 sources are averagely more luminous than the sources in \citet{Ho97}. However, we find our sources are relatively weaker in high luminous end than the sources in \citet{Stern12}. The different properties of these source groups are caused by the different populations. \citet{Stern12} select their sample from a population that is detected broad H$\alpha$ and while the sources in \citet{Ho97} are selected from a population that has relatively low luminous. Our sources are picked up from a Seyfert 2 sample that is  relatively low luminous in AGNs and our sources are more similar to that in \citet{Ho97}.

\begin{table}
\caption{FITS Format}
\label{tab:fits_format}
\centering
\begin{tabular}{ccc}
\toprule
Column  & NAME & Description \\
\hline
1 & R.A.  & Right ascension (J2000) in decimal degrees  \\
2 & DEC. & Declination (J2000) in decimal degrees    \\
3 & z & Redshift  \\
4 & plate & Spectroscopic plate number in SDSS \\
5 & MJD & Number of spectroscopic Modified Julian date (MJD) in SDSS \\
6 & FiberID & Spectroscopic FiberID number in SDSS \\
7 & L$_{\mathrm{H\alpha, broad}}$  & Luminosity of broad H$\alpha$ component [ergs/s] \\
8 & $\sigma_{\mathrm{L_{H\alpha, broad}}}$  & Uncertainty in L$_{\mathrm{H\alpha, broad}}$ \\
9 & L$_{\mathrm{H\alpha, narrow}}$  & Luminosity of narrow H$\alpha$ component [ergs/s] \\
10 & $\sigma_{\mathrm{L_{H\alpha, narrow}}}$  & Uncertainty in L$_{\mathrm{H\alpha, narrow}}$ \\
11 & FWHM$_{\mathrm{H\alpha, broad}}$  & FWHM of broad H$\alpha$ component [km/s] \\
12 & $\sigma_{\mathrm{FWHM_{H\alpha, broad}}}$  & Uncertainty in FWHM$_{\mathrm{H\alpha, broad}}$ \\
\toprule
\end{tabular}
\end{table}

In this work, we identify a total of 1297 Seyfert 1.9 galaxies and 157 intermediate Seyfert (Sy1.2, Sy1.5, and Sy1.8) sources that have a broad H$\beta$ component. We present the measurement of the H$\alpha$ emission line of the 1297 Seyfert 1.9 galaxies in an online catalog\footnote{\url{openuniverse.asi.it/syfert1.9}} and the catalog format is described in  Table~\ref{tab:fits_format}. The complicated spectra of the intermediate Seyfert sources are hard to be picked out from the amount of observation data. A simple method was to use flux ratio in the early stage \citep{Whittle92,Winkler92}. This method is quick for dealing with amount observation data. However, a flux ratio does not reflect the physic meaning of line profile and this method did not assign any ratio to Seyfert 1.9 galaxies. Recently, a common way to pick out intermediate Seyferts is by fitting candidate spectra to check if the emission lines exist the second component or estimating excess flux near the emission line due to a wing structure \citep{Wang08,Stern12}. The fitting methods usually spend a lot of time on fitting candidate spectra. In our method, we build a 1D CNN model and train this model with a few Seyfert 1.9 sources. The training process takes few hours and the test process for tens of thousands of sources only takes few minutes. The advantage of our method is that we do not have to spend a lot of time fitting lots of candidate spectra. We only focus on fitting the target spectra that are selected from our CNN model. The other advantage is that the model can reduce contamination of Seyfert 1.9 in the Seyfert 2 sample.

\begin{figure}[ht]
\begin{center}
\includegraphics[scale=0.65,angle=0]{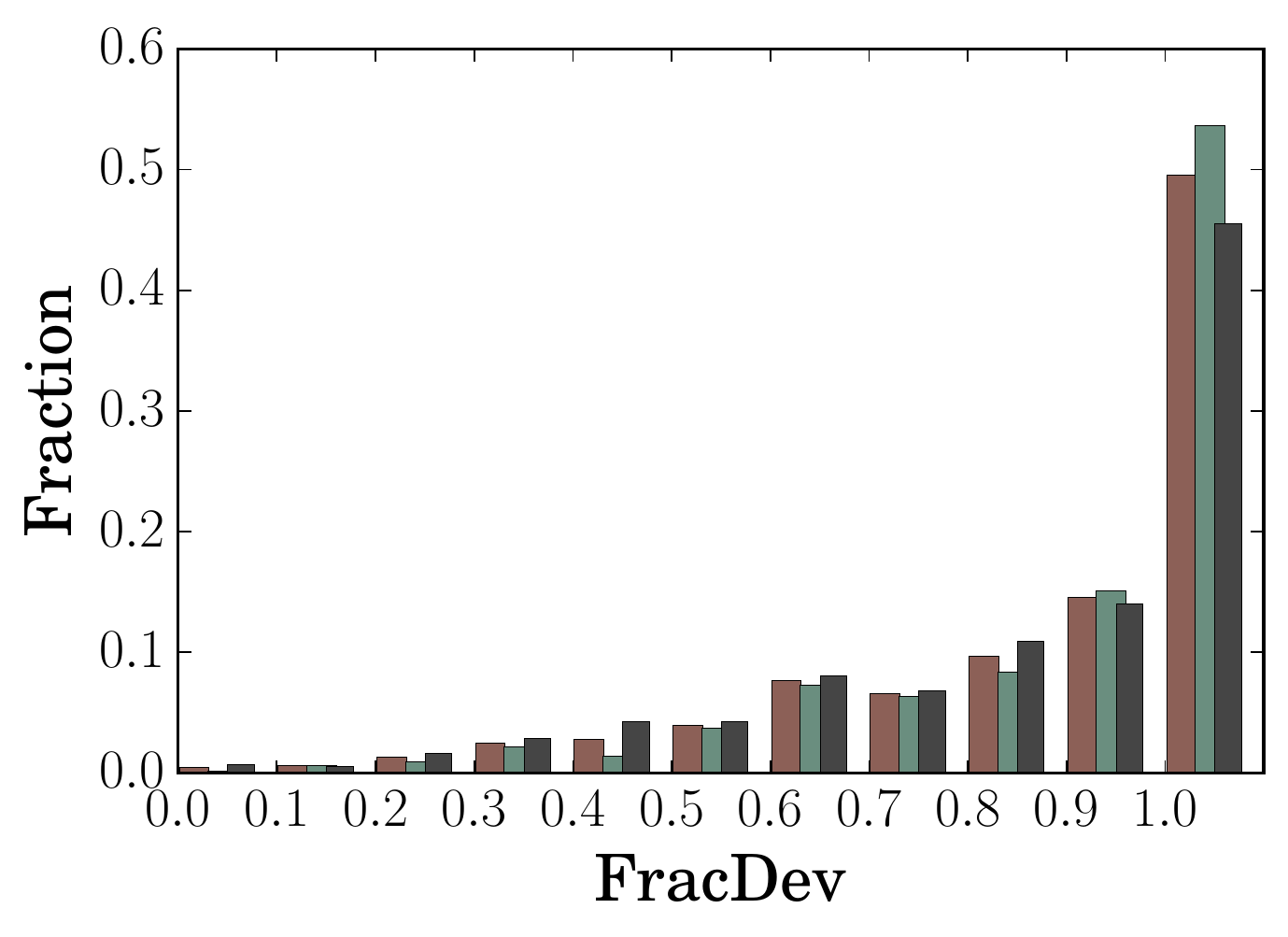}
\caption{{\it FracDev} distributions. Brown represents the 1297 Seyfert 1.9 galaxies. Green represents the  human-selected Sy1.9 sample. Dark-gray represents machine-selected Sy1.9 sample. }
\label{fig:fracd}
\end{center}
\end{figure}

{\it FracDev} is an indicator of host galaxy morphology. This parameter describes the bulge contribution in galaxies and ranges from 0 to 1. A bulge-dominated galaxy has {\it FracDev} value close to 1. We show the {\it FracDev} distribution of our Seyfert 1.9 sample in Fig.~\ref{fig:fracd}. We find both of our Seyfert 1.9 samples show bulge dominant distribution. The percentage of {\it FracDev} =1 are 53.7\% and 45.6\% for human-selected Sy1.9 and machine-selected Sy1.9 sample. We estimate a K-S test for these two Seyfert 1.9 samples and obtain statistic D = 0.0967 with a p-value = 0.0042, indicating that the two distributions have a low probability to be drawn from the same population. The total {\it FracDev} distribution of the two Seyfert 1.9 samples is also dominated by {\it FracDev} =1 and the percentage of {\it FracDev} =1 is 49.6\%. We compare the {\it FracDev} distribution of Seyfert 1.9 to that of Seyfert 1 and Seyfert 2 galaxies in \citet{Chen17}. We find both of our Seyfert 1.9 samples and the Seyfert 1 show bulge dominant distribution. We estimate a K-S test for 1297 Seyfert 1.9 and Seyfert 1 and obtain statistic D = 0.0602 with a p-value = 0.0014. This result indicates that Seyfert 1.9 galaxies and Seyfert 1 galaxies have a low probability to be drawn from the same population. On the other hand, the {\it FracDev} distribution of Seyfert 1.9 galaxies is different from that of Seyfert 2 galaxies. We estimate a K-S test for 1297 Seyfert 1.9 and Seyfert 2 and the test result is statistic D = 0.2545 with a p-value = 8.56965 $\times$ 10$^{-54}$. This result indicates that Seyfert 1.9 galaxies have an extremely low probability to have the same population as Seyfert 2 galaxies. These results suggests that the different types of the Seyferts might be related to its host galaxy morphology \citep{Chen17}.


\begin{figure}[ht]
\begin{center}
\includegraphics[scale=0.3,angle=0]{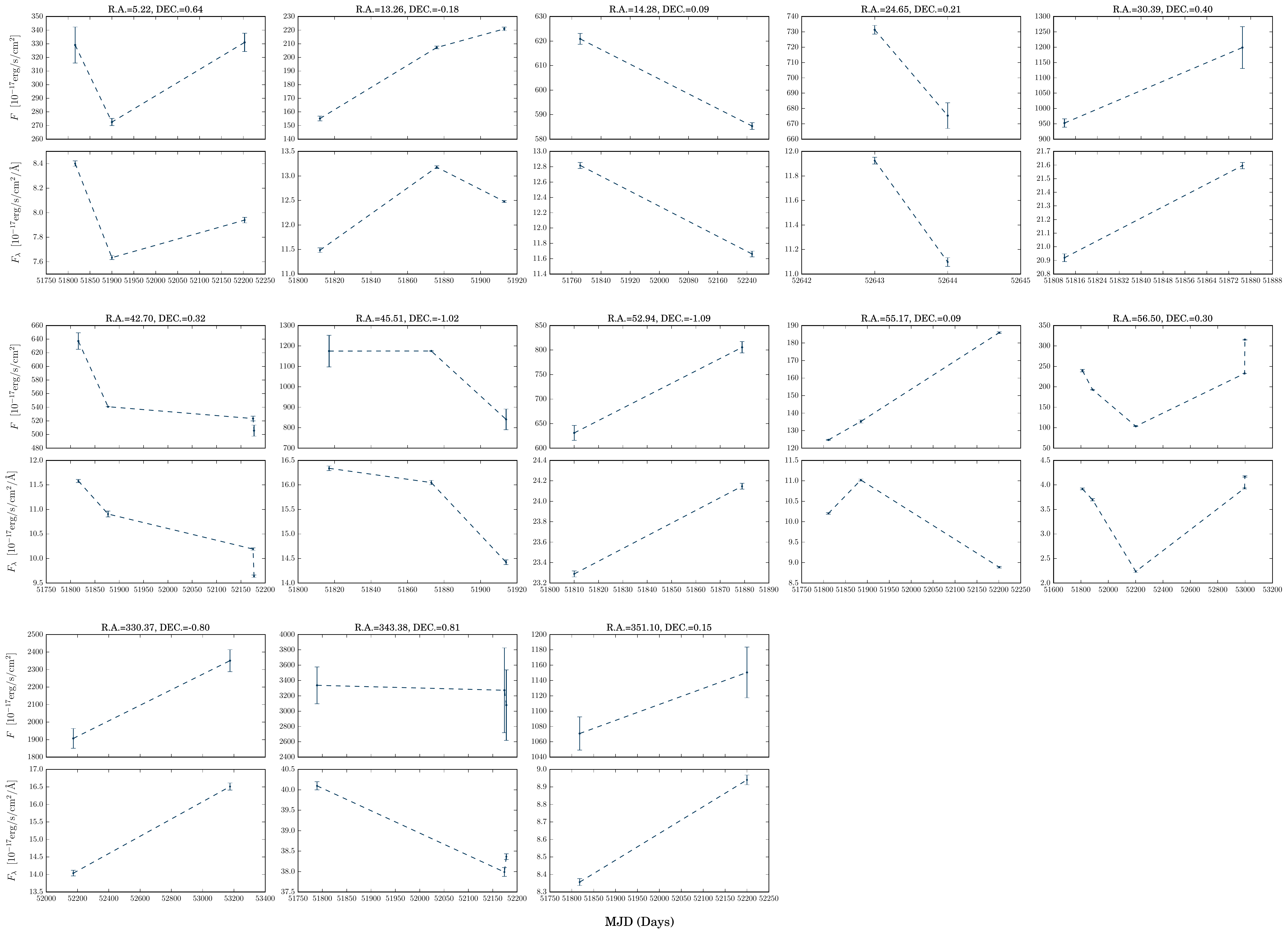}
\caption{Light curve of the 13 Seyfert 1.9 galaxies. In each figure, the top panel is the flux of broad H$\alpha$ component and the bottom panel is flux density of continuum level at H$\alpha$.}
\label{fig:lc_all}
\end{center}
\end{figure}

We search for repeatable spectra of our 1297 Seyfert 1.9 sources in the Stripe 82 region \citep{Annis14}. However, we only have 13 sources with multiple spectra and obtain the spectra from Catalog Archive Server (CAS) Stripe 82 database. We list coordinates and detailed observation information of the 13 sources in Table \ref{tab:sources}. We fit the spectra with a linear function as a pseudo continuum and two Gaussian components as a broad and narrow component and the detail fitting process is described in Appendix \ref{appendix:a}. We derive flux of broad H$\alpha$ component and flux density of continuum level at H$\alpha$, and the flux uncertainty is estimated from fitting results of the least square. We show the light curves of the 13 sources in Fig.~\ref{fig:lc_all}. We find that the light curves of our 13 Seyfert 1.9 galaxies show variability in broad H$\alpha$ component and continuum level.  In most light curves, the flux of broad H$\alpha$ component varies with continuum level. This suggests that the broad H$\alpha$ emission is related to photoionization from galaxy center \citep{Peterson93}. However, we find two sources do not show  simultaneous changes in the flux of broad H$\alpha$ component and continuum level. The light curve period of the first source (R.A.=13.26 and DEC.=-0.18) is from MJD= 51876 to 51913 and the light curve period of the second source (R.A.=55.17 and DEC.=0.09) is from MJD=51885 to 52201. The nonsimultaneous variation might be caused by a time delay between central ionization and BLR cloud and the delay time scale ranges from few days to few weeks. We find the first source has a time scale of 37 days and conclude that this source has more possibility to be caused by the effect of the time delay. The second source has a time scale of 316 days and the time scale is much longer than few weeks. This might suggest that this source has a relatively large BLR. However, we find the continuum level decreases from $11.02\times10^{-7}$ to $8.88\times10^{-7}$ [ergs/s/cm$^2$/\AA] and the variation is $\approx 19.3\%$. The insignificant variation of the continuum level is difficult for us to explain and conclude the possible reason behind the second source.

\begin{table}
\caption{Coordinates, redshifts, observation information for our 13 sources}
\label{tab:sources}
\centering
\begin{tabular}{lcccccc}
\toprule
 & R.A. & Dec. & z & plate & MJD & fiberID  \\
\hline
 & 52.935422 & -1.0879826 & 0.0851696 & 415 & 51810 & 246  \\
 & & & 0.085141 & 415 & 51879 & 248  \\
 & 30.385701 & 0.39812053 & 0.078188 & 404 & 51877 & 347 \\
 & & & 0.078162 & 404 & 51812 & 343 \\
 & 13.262922 & -0.17965537 & 0.13831 & 394 & 51913 & 161  \\
 & &  & 0.138298 & 394 & 51812 & 176  \\
  & &  & 0.138837 & 394 & 51876 & 166  \\
 & 5.2221571 & 0.63686113 & 0.144553 & 390 & 51900 & 456 \\
 & &  & 0.144353 & 390 & 51816 & 454 \\
 & &  & 0.144547 & 688 & 52203 & 422 \\
 & 42.700671 & 0.31673936 & 0.18769 & 707 & 52177 & 631  \\
 & & & 0.187389 & 410 & 51816 & 360  \\  
 & & & 0.187546 & 410 & 51877 & 351  \\ 
 & & & 0.187462 & 708 & 52175 & 393  \\ 
 & 55.166922 & 0.088592222 & 0.130904 & 714 & 52201 & 496 \\
 & & & 0.130917 & 416 & 51811 & 479 \\
 & & & 0.130941 & 416 & 51885 & 463 \\
 & 343.38084 & 0.80708478 & 0.0724755 & 676 & 52178 & 453 \\
 & & & 0.072455 & 379 & 51789 & 507 \\
 & & & 0.072478 & 676 & 52174 & 458 \\
 & 351.10088 & 0.14853945 & 0.148683 & 383 & 51818 & 493 \\
 & & & 0.148586 & 680 & 52200 & 514 \\
 & 14.282579 & 0.088951841 & 0.195946 & 395 & 51783 & 393  \\
 & & & 0.19587 & 693 & 52254 & 347  \\
 & 45.505091 & -1.0216328 & 0.166582 & 411 & 51817 & 260  \\
 & & & 0.166661 & 411 & 51873 & 248  \\
 & & & 0.166662 & 411 & 51914 & 253  \\
 & 330.36646 & -0.80064708 & 0.110172 & 1032 & 53175 & 54  \\
 & & & 0.110156 & 372 & 52173 & 56  \\
 & 24.6492 & 0.20733 & 0.124992 & 1078 & 52643 & 571  \\
 & & & 0.125044 & 1077 & 52644 & 358  \\
 & 56.497262 & 0.303874 & 0.161065 & 1632 & 52996 & 588  \\
 & & & 0.16106 & 416 & 51811 & 624  \\
 & & & 0.161073 & 416 & 51885 & 624  \\
 & & & 0.161196 & 714 & 52201 & 629  \\
 & & & 0.161058 & 1633 & 52998 & 394  \\
\toprule
\end{tabular}
\end{table}

\section{Summary}

We build a 1D CNN model to collect Seyfert 1.9 sources from Seyfert 2 sample. The model trained by intermediate Seyfert or Seyfert 1.9 galaxies has more than 80\% precision for classifying correctly Seyfert 1.9 galaxies. Besides, the model provides a new Seyfert 1.9 sample which was missed in visual inspection. We also use the new Seyfert 1.9 sample as training sample to improves the performance of our model and obtain a 91\% precision for classifying correctly Seyfert 1.9 galaxies.

In this work, we have original human-selected Sy1.9 sources that are picked out by visual inspection and  machine-selected Sy1.9 sources that are picked out by our CNN model. In total, we identify 1297 Seyfert 1.9 galaxies and decompose their H$\alpha$ emission line by fitting 2 Gaussian components. We find the two Seyfert 1.9 samples have a similar distribution of their FWHM$_\mathrm{H\alpha}$ and H$\alpha$ luminosity. However, we find the machine-selected Sy1.9 sample has slightly higher FWHM$_\mathrm{H\alpha}$ and lower luminous than the human-selected Sy1.9 sample. This suggests that our model picks out the relatively weak Seyfert 1.9 sources that are usually missed by visual inspection. We check the properties of the BPT diagram of our Seyfert 1.9 samples and find the two Seyfert 1.9 samples have a similar distribution in the BPT diagram. We also compare the BPT diagram of the Seyfert 1.9 samples to that of the Seyfert 2 sample and find the Seyfert 1.9 samples distribute with relatively higher [NII]/H$\alpha$ and weaker [OIII]/H$\beta$ than Seyfert 2 galaxies in BPT diagram. From the distribution of the host galaxy morphology of the Seyfert 1.9 sample, we find our Seyfert 1.9 galaxies are dominated by galaxies with high {\it FracDev} value and the distributions of host galaxy morphology of the Seyfert 1.9 are more similar to that of Seyfert 1 galaxies instead that of Seyfert 2 galaxies. This suggests that the difference between different types of Seyfert galaxies might be related to the host galaxy morphology. Finally, we provide an online catalog of our 1297 Seyfert 1.9 galaxies with the measurement of the H$\alpha$ emission line.


\begin{acknowledgments}

I thank anonymous referee for helpful comments.\\
I thank P.C. Yu, Y. Wang, P. Giommi for discussion and useful comments.\\
I especially thank Paolo Giommi and SSDC people for building an online page for Seyfert 1.9 catalog.\\

Funding for SDSS-III has been provided by the Alfred P. Sloan Foundation, the Participating Institutions, the National Science Foundation, and the U.S. Department of Energy Office of Science. The SDSS-III web site is http://www.sdss3.org/. SDSS-III is managed by the Astrophysical Research Consortium for the Participating Institutions of the SDSS-III Collaboration including the University of Arizona, the Brazilian Participation Group, Brookhaven National Laboratory, Carnegie Mellon University, University of Florida, the French Participation Group, the German Participation Group, Harvard University, the Instituto de Astrofisica de Canarias, the Michigan State/Notre Dame/JINA Participation Group, Johns Hopkins University, Lawrence Berkeley National Laboratory, Max Planck Institute for Astrophysics, Max Planck Institute for Extraterrestrial Physics, New Mexico State University, New York University, Ohio State University, Pennsylvania State University, University of Portsmouth, Princeton University, the Spanish Participation Group, University of Tokyo, University of Utah, Vanderbilt University, University of Virginia, University of Washington, and Yale University.

\end{acknowledgments}

\appendix

\section{Fitting H$\alpha$ emission line}
\label{appendix:a}

The H$\alpha$ emission line profile of the Seyfert 1.9 galaxies shows a strong narrow component superimpose a weak broad component. We fit two Gaussian components to the H$\alpha$ emission line of the Seyfert 1.9 galaxies. First, we mask all emission lines in the region of 6450\AA~to 6700\AA~to find the continuum level. We fit a linear component as the continuum part by the least square method. Second, we only mask the region of [NII]$\lambda$6548 and [NII]$\lambda$6583 emission lines and fit 2 Gaussian components with a linear component which is derived in early step and will be fixed in this fitting by Levenberg-Marquardt algorithm. After deriving the parameters of the 2 Gaussian, we estimate the full-width at half-maximum (FWHM) of the two components and obtain the H$\alpha$ line flux and luminosity of the two components from fitting results.


\bibliography{sample631}{}
\bibliographystyle{aasjournal}



\end{document}